\documentclass[12pt,a4paper]{article}
\usepackage[utf8]{inputenc}
\usepackage[total={6.0in, 8.7in}]{geometry}
\title{\Large\textbf{Dyons with phase $\delta_{\theta}=n\theta$}}
\author{Ricardo Heras\thanks{e-mail: ricardo.heras@ou.ac.uk}}
\date{{\small \textit{School of Physical Sciences, The Open University,\\ Walton Hall, Milton Keynes MK7 6AA, UK}}}

\usepackage{hyperref}

\usepackage{bm}
\usepackage{graphicx}
\usepackage[mathscr]{euscript}
\usepackage{amsmath}
\usepackage{enumitem}
\usepackage{amsfonts}
\usepackage{braket}

\hypersetup{
  colorlinks = true,
  citecolor = blue,
  urlcolor = blue
}

\newcommand{\gradv}{\boldsymbol{\nabla}}
\def\v#1{{\bf#1}}

\begin{document}
\maketitle

\begin{abstract}
\noindent In a recent paper  (Heras in Eur. Phys. J. Plus 138:329, 2023), we have demonstrated that when a dyon encircles an infinitely long solenoid enclosing electric and magnetic fluxes, its wave function accumulates a quantum phase invariant under electromagnetic duality transformations. In this paper, we show that this phase, in conjunction with the Witten effect, gives rise to a topological phase proportional to the vacuum angle $\theta$ and thereby connected with CP violation. We show that this phase becomes quantised in a vacuum state $\delta_{\theta}=n\theta$ and that the most general vacuum state associated with this quantisation identifies with an Abelian form of the  $\theta$-vacua. We discuss two hypothetical interference effects in the vacuum where the angle $\theta$ could manifest.
\end{abstract}

\vskip 9pt
\noindent Dyons are hypothetical particles having electric and magnetic charges. These particles hold significant appeal as they offer an explanation for the quantisation of electric charge \cite{1,2,3} and are a prediction of grand unified theories \cite{4}. The concept of dyons was initially put forward in 1968 by Schwinger \cite{1} and by Zwanziger \cite{2}, both of who derived the Schwinger-Zwanziger (SZ) quantisation condition
\begin{align}
q_1g_2 - q_2 g_1 = N \frac{\hbar c}{2},
\end{align}
which holds for any pair of dyons with electric and magnetic charges ($q_1,g_1$) and ($q_2,g_2$), where $N$ is an integer and Gaussian units are adopted. In 1979, Witten \cite{5} noted that Eq.~(1), was satisfied not only by the standard quantisation rules for electric and magnetic charges: $q=n_qe$ and $g=n_g g_{0},$ where $n_q$ and $n_g$ are integers, $e$ is the electron charge strength, and $g_{0}=e/(2\alpha)$ is the quantum of magnetic charge, with $\alpha$ denoting the fine structure constant, but also by the quantisation rules
\begin{align}
q=n_q e +n_g\frac{e\theta}{2\pi},  \quad g=n_g g_{0},
\end{align}
where $\theta$ is the vacuum angle, indicating that Eq.~(2) hold for theories violating CP symmetry. The $\theta$-term in Eq.~(2) characterises the Witten effect and has played an influential role in supersymmetric gauge theories \cite{6,7,8}. While Witten derived the quantisation rules in Eq.~(2) within the framework of non-Abelian gauge theories, Coleman \cite{9} subsequently demonstrated that these rules apply also to Abelian gauge theories, which will be our primary interest in this paper. More specifically, we will examine the implications of the Witten effect on the recently discussed quantum phase \cite{10} 
\begin{align}
\delta_{D}= \frac{n}{\hbar c}(q\Phi_{\text{m}} - g \Phi_{\text{e}}),
\end{align}
that accumulates the wave function of a dyon of charges $q$ and $g$ upon encircling an infinitely long solenoid enclosing uniform magnetic and electric fluxes $\Phi_{\text{m}}$ and $\Phi_{\text{e}}$. We shall refer to this solenoid as a dual flux tube and Eq.~(3) as the dyon phase. This phase is topological because it depends on a winding number $n$ characterising the number of times the dyon encircles the dual flux tube and reflects a nonlocal interaction since the enclosed fluxes act on the dyon in a region where they vanish. The dyon phase is also invariant under electromagnetic duality transformations providing a unified model of the Aharonov-Bohm (AB) phase \cite{11} with its corresponding dual-phase \cite{12,13} (see also the related work \cite{14}). Moreover, the dyon phase can be used to heuristically derive the SZ quantisation condition\footnote{If we envision a dyon of charges $(q_2,g_2)$ as the end of a semi-infinite dual  Dirac string localised along the negative $z$-axis, then this configuration has the associated fields $\v B = \v B_{\rm d} + \v B_{\rm s}$ and $\v E = \v E_{\rm d} + \v E_{\rm s}$, where $\v B_{\rm d}= g_2\hat{\v r}/r^2$ and $\v E_{\rm d}= q_2\hat{\v r}/r^2$ denote the fields of the dyon, while $\v B_{\rm s}=4\pi g_2\delta(x)\delta(y)\Theta(-z)\hat{\v z}$ and $\v E_{\rm s}=4\pi q_2\delta(x)\delta(y)\Theta(-z)\hat{\v z}$ denote the fields of the semi-infinite string attached to the dyon, where $\Theta$ is the Heaviside step function. The fields of the string have the corresponding fluxes $\Phi^{\rm s}_{\rm m}=\int_{S_{\varepsilon}}\v B_{\rm s}\cdot d \v S= 4\pi g_2 \Theta(-z)$ and $\Phi^{\rm s}_{\rm e}=\int_{S_{\varepsilon}}\v E_{\rm s}\cdot d \v S=4\pi q_2 \Theta(-z)$, where $S_{\varepsilon}$ is the infinitesimal surface pierced by the string. Now, if a dyon of charges $(q_1,g_1)$ encircles the string once, it appears reasonable to argue that its wave function would acquire the dyon phase $\delta_D =(q_1\Phi^{\rm s}_{\rm m}-g_1\Phi^{\rm s}_{\rm e})/(\hbar c) =\Theta(-z)4\pi(q_1 g_2 - q_2 g_1)/(\hbar c)=4\pi(q_1 g_2 - q_2 g_1)/(\hbar c)$, where $\Theta(-z)=1$ because the encircling dyon lies at at $z<0$. By requiring the string to be ``unobservable," we impose a trivial value for the dyon phase $\delta_D=2\pi N$, leading to the SZ quantisation condition. However, this argument is heuristic and subject to criticism for the following reasons: (i) it is not possible to define a set of \textit{global} vector potentials $\v A_{\rm s}$ and $\v C_{\rm s}$ associated with the semi-infinite string fields $\v B_{\rm s}$ and $\v E_{\rm s}$ (this can only be accomplished with an infinite string), (ii) setting the dyon phase to a trivial value $\delta_D=2\pi N$ does not imply the string is unobservable: its interference effects would be undetected, but the string would still be observable through the Lorentz force and the electromagnetic angular momentum since the string fields $\v B_{\rm s}$ and $\v E_{\rm s}$ are non-vanishing, and (iii) considering that a \textit{semi-infinite} string and a \textit{infinite} string are \textit{not} homeomorphic (topologically equivalent), it cannot be assumed a priori that the phase $\delta_D$ predicted with an infinite string \cite{10} also arises with a semi-infinite string.} in a manner akin the AB phase is used to obtain the Dirac quantisation condition \cite{15}.

Consider first the standard quantisation rules $q=n_qe$ and $g=n_g g_{0}$ in the dyon phase to obtain 
\begin{align}
\delta_{D} = \frac{n}{\hbar c}(n_q e\Phi_{\text{m}} - n_g g_{0} \Phi_{\text{e}}).
\end{align}
Interestingly, this phase vanishes $\delta_D=0$ when considering a dyon with elementary charges $(n_q= n_g=1$) and when the fluxes through the dual flux tube are given by the magnetic and electric flux quanta $(\Phi^{0}_{\rm m}= 2\pi \hbar c/e$, $\Phi^{0}_{\rm e}=2\pi \hbar c/g_{0})$. We will subsequently provide an explanation for this result. 

We now consider the Witten effect on the dyon phase. In this case, we use Eq.~(2) and obtain
\begin{align}
\delta_{D} = \frac{n}{\hbar c}(n_q e\Phi_{\text{m}} - n_g g_{0} \Phi_{\text{e}})+ \delta_{\theta},
\end{align}
where the  $\theta$-phase is defined by
\begin{align}
\delta_{\theta}= nn_g\frac{\theta e  \Phi_{\rm m}}{2\pi \hbar c}.
\end{align}
Since this phase depends on the vacuum angle $\theta$, then we can say that this phase is a manifestation of CP violation. Indeed, this is the case except for two mutually excluding values of the angle $\theta$ in which CP is conserved \cite{5}: $\theta=0$ and $\theta=\pi$. The former value implies that dyons follow the integer electric charge quantisation $q=n_qe$, whereas the latter that dyons follow the fractional electric charge quantisation $q=n_qe + n_ge/2$. We have obtained Eq.~(4) by assuming the standard quantisation rules $q=n_qe$ and $g=n_g g_{0}$ and this means we have tacitly assumed $\theta =0$ when CP is conserved, which implies: $\delta_{\theta}=0$ (we will discuss the case $\theta=\pi$ at the end of this paper). We also note that when we apply the transformation $\theta\to\theta + 2\pi$ to the dyon phase in Eq.~(5), it preserves its form because this transformation shifts its electric charge number $n_q\to n_q + n_g$. Accordingly, the dyon phase is periodic in $\theta$ and therefore the transformation $\theta\to\theta + 2\pi$ is a symmetry of this phase, which is consistent with the fact that relevant physical quantities should be periodic in the angle $\theta$ \cite{9}.

We can alternatively express the $\theta$-phase in Eq.~(6) using the relation $n_g=2 \alpha g/e$, which follows from the quantisation rule $g=n_g g_{0}$ with $g_{0} = e/(2 \alpha)$. The alternative form reads
\begin{align}
\delta_{\theta}= n\frac{\alpha \theta g\Phi_{\rm m}}{\pi \hbar c},
\end{align}
which shows that the $\theta$-phase is topological because it depends on the winding number $n$ and is nonlocal because the magnetic charge $g$ of the dyon is affected by the magnetic flux $\Phi_{\rm m}$ in a region where this flux is excluded. In particular, if $g=g_{0}$ and $\Phi_{\rm m}\!=\!\Phi^{0}_{\rm m}\!=\! 2\pi \hbar c/e$ then the  $\theta$-phase becomes
\begin{align}
\delta_{\theta}= n \theta.
\end{align}
This particular form of the $\theta$-phase describes a topological quantisation ($n$ is a winding number) of the vacuum angle $\theta$ (under the assumption that $\theta$ is a universal constant \cite{16}). When $n=1$, it follows that $\theta=\delta_{\theta}$ and this means that the angle $\theta$ may be interpreted as a topological phase whose origin is the dyon phase given by Eq.~(5). We also note that Eq.~(8) can alternatively be obtained from Eq.~(5) when the dyon has the elementary electric and magnetic charges $(n_q= n_g=1$) and the fluxes through the dual flux tube are given by the flux quanta $(\Phi^{0}_{\rm m}= 2\pi \hbar c/e$ and $\Phi^{0}_{\rm e}=2\pi \hbar c/g_{0}$).

A Hamiltonian treatment allows us to see that the topological quantisation of the $\theta$-phase in Eq.~(8) corresponds to a vacuum state. The Hamiltonian of the system formed by a non-relativistic dyon encircling the dual flux tube is given by \cite{10}
\begin{align}
\widehat{H} = \frac{1}{2m}\bigg(-i\hbar\gradv- \frac{(q \v A + g \v C)}{c}  \bigg)^2 + V,
\end{align}
where $m$ is the mass of the dyon, $V$ is a scalar potential associated with a mechanical force that keeps the dyon encircling the dual flux tube, $\v A$ is the magnetic vector potential associated with the magnetic flux $\Phi_{\rm m}$, and $\v C$ is the electric vector potential associated with the electric flux $\Phi_{\rm e}$. Using Eq.~(9), we obtain the time-independent Schr\"odinger equation: $\widehat{H}\ket{\psi}\!=\!E \ket{\psi}$, where $\ket{\psi}$ is the state of the dyon with energy $E$. Since the fields outside the dual flux tube vanish $\v B = \gradv \times \v A=0$ and $\v E = -\gradv \times \v C=0$, then the potentials can be expressed as \cite{10,17}: $\v A = \gradv \chi$ and $\v C=\gradv\xi,$ where $\chi = \phi\, \Phi_{\rm m}/(2\pi)$ and $\xi=-\phi\,\Phi_{\rm e}/(2\pi)$ are multi-valued functions of the azimuthal coordinate $\phi$. Using these results and following the procedure given in Ref.~\cite{10}, we obtain the corresponding solution
\begin{align}
\ket{\psi}\!=\! {\rm e}^{in(q\Phi_{\rm m} - g \Phi_{\rm e})/(\hbar c)}{\rm e}^{i\int_{\gamma}[(q \v A + g \v C)/(\hbar c)]\cdot d \v x'}\!\ket{\psi_{0}},
\end{align}
where $\ket{\psi_{0}}$ is the state when $\v A=0$ and $\v C=0$ in Eq.~(9). The first phase in Eq.~(10) is the dyon phase $\delta_{D}$, which deals with the number of times the dyon encircles the dual flux tube. The second phase is a local phase accounting for the open trajectory $\gamma$ the dyon takes before completing another turn around the dual flux tube. We note that the states $\ket{\psi}$ and $\ket{\psi_{0}}$ have a winding number dependency: $\ket{\psi}=\ket{\psi(n)}$ and $\ket{\psi_{0}}=\ket{\psi_{0}(n)}$ because both states are defined in the non-simply connected space: $\mathbb{R}^3 - \{S^1\times \mathbb{R}\}$ (the Euclidean space minus the infinite cylinder).

The minimum value of the Hamiltonian corresponds to a vacuum state \cite{18,19}. In our case, the vacuum state depends on whether or not CP is violated. We note that if $V=0$, then the dyon would cease to encircle the dual flux tube. Therefore, we set $V=V_{0},$ where $V_{0}$ represents minimum potential required to maintain the motion of the dyon around the dual flux tube. The identification of the potential $V=V_{0}$ remains valid irrespective of whether CP is conserved or not.

Consider first the case in which CP is conserved. If we assume a dyon with the elementary charges $q=e$ and $g=g_{0}$ and consider the flux quanta $\Phi^{0}_{\rm m}=2\pi \hbar c/e$ and $\Phi^{0}_{\rm e}=2\pi \hbar c/g_{0}$, then we obtain the relation $e\Phi^{0}_{\rm m}- g_{0}\Phi^{0}_{\rm e}=0$, which yields a vanishing conjugate momentum $(q \v A + g \v C)/c=0$. Considering this result together with  $V=V_{0}$ in Eq.~(9), we obtain the minimised Hamiltonian $\widehat{H}= -\hbar^2\gradv^2/(2m) + V_{0}$ and the state in Eq.~(10) reduces to $\ket{\psi}=\ket{\psi_{0}}$. This shows that in a CP-conserving vacuum, we have the vanishing of the dyon phase $\delta_D=0$ and this explains our earlier observation as to why this phase vanishes when considering the elementary charges and the flux quanta. We will now consider the case in which CP is not conserved. Equation~(2) with $n_q= n_g=1$ yields a dyon with the elementary charges $q=e + e\theta/(2\pi)$ and $g=g_{0}$. These values together with the flux quanta $\Phi^{0}_{\rm m}=2\pi \hbar c/e$ and $\Phi^{0}_{\rm e}=2\pi \hbar c/g_{0}$ yield the conjugate momentum
\begin{align}
 \frac{(q\v A + g\v C)}{c} = \hbar\gradv \beta,
\end{align}
where $\beta\!=\!\theta \phi/(2\pi)$ is a multi-valued function of the azimuthal coordinate. Using Eq.~(11) together with $V\!=\!V_{0}$ in Eq.~(9), we obtain the minimised Hamiltonian for the case in which CP is violated
\begin{align}
\widehat{H}= \frac{\hbar^2}{2m}\bigg(i \gradv + \gradv \beta \bigg)^2 + V_{0}.
\end{align}
The corresponding state of the dyon follows from Eq.~(10), which takes the specific form
\begin{align}
\ket{\psi}={\rm e}^{i n(e \Phi^{0}_{\rm m}-g_{0}\Phi^{0}_{\rm e})/(\hbar c)}{\rm e}^{in \theta e \Phi^{0}_{\rm m}/(2\pi \hbar c)}{\rm e}^{i\int_{\gamma}\gradv'\!\beta\cdot d \v x'}\ket{\psi_{0}}={\rm e}^{in\theta}{\rm e}^{i\phi \theta/(2\pi)}\ket{\psi_{0}},
\end{align}
where we have used the results $e\Phi^{0}_{\rm m}- g_{0}\Phi^{0}_{\rm e}=0$, $e\Phi^{0}_{\rm m}/(2\pi \hbar c)=1$, and $\int_{\gamma}\gradv'\beta\cdot d \v x'=\phi \theta/(2\pi)$. For convenience, we adopt the notation $\ket{n}={\rm e}^{i\phi \theta/(2\pi)}\ket{\psi_{0}},$ where $\ket{n}$ is a state that depends on the winding number $n$ since $\ket{\psi_{0}}=\ket{\psi_{0}(n)}$. In terms of $\ket{n}$, the state of the dyon in Eq.~(13) becomes
\begin{align}
\ket{\psi}={\rm e}^{i n \theta} \ket{n}.
\end{align}
This equation describes a vacuum state characterised by the definite winding number $n$. We clearly identify in Eq.~(14) the quantised $\theta$-phase: $\delta_{\theta}=n\theta$. Accordingly, we have shown that when the dyon is in a CP-violating vacuum, the dyon phase in Eq.~(5) reduces to the quantised $\theta$-phase in Eq.~(8).

We note that Eq.~(14) does not represent the most general vacuum state. If the dyon takes a further turn around the dual flux tube, then $\phi \to \phi + 2\pi$ and its state changes to $\ket{\psi}={\rm e}^{i (n +1)\theta} \ket{n+1}$, or equivalently, $\ket{\psi}={\rm e}^{i n'\theta} \ket{n'},$ which now describes a vacuum state characterised by the definite winding number $n'=n+1$. Since $n\neq n'$, then the states $\ket{\psi}={\rm e}^{i n \theta} \ket{n}$ and $\ket{\psi}={\rm e}^{i n' \theta} \ket{n'}$ correspond to different vacua. This method implies that there are an infinite number of equivalent vacua corresponding to different winding numbers. However, we would expect that the most general vacuum state should be invariant under dyon rotations. We can solve this problem by noting that the state $\ket{n}$ forms a complete basis, and therefore we can obtain the superposition
\begin{align}
\ket{\theta}= \sum^{+\infty}_{n=-\infty} {\rm e}^{i n \theta} \ket{n},
\end{align}
which is form invariant under the transformation  $\phi\to \phi + 2\pi$ modulo a constant phase factor. Therefore, Eq.~(15) represents the most general vacuum state of the dyon. Interestingly, Eq.~(15) has the form of the $\theta$-vacua of non-Abelian gauge theories \cite{20} (see also \cite{18,21}). However, Eq.~(15) is Abelian since the potentials in Eq.~(11) are $U(1)$ gauge fields, and also $\ket{\theta}$ is a state function and not a state functional as in the non-Abelian $\theta$-vacua \cite{20}. The fact that the Abelian $\theta$-vacua in Eq.~(15) and the non-Abelian $\theta$-vacua have the same form illustrates the topological nature of the CP-violating vacuum. We believe that this is because the Witten effect is topological and therefore manifests itself in Abelian and non-Abelian theories \cite{9}. We also note that Wilczek \cite{22} suggested that dyons in non-Abelian theories should have a vacuum structure similar to Eq.~(15).

We will now briefly discuss two hypothetical vacuum interference effects, where the angle $\theta$ could manifest. Consider first a two-slit interference experiment in which a dyon propagates from a source passes through one of the two slits on a first screen and is detected on a second screen. If we insert a dual flux tube between the screens, then the normalised state of the dyon is given by a superposition of states going to the left and right of the dual flux tube:  $\ket{\psi}= (\ket{\psi_L} + \ket{\psi_R})/\sqrt{2}$. The states $\ket{\psi_L}$ and $\ket{\psi_R}$ follow from Eq.~(13) when we set $n\!=\!0$ since none of these states encircle the dual flux tube separately. But since the difference in the trajectories $\gamma_L$ and $\gamma_R$, associated with the states $\ket{\psi_L}$ and $\ket{\psi_R}$, forms a loop $C\!=\!\gamma_{R}-\gamma_{L}$ enclosing the dual flux tube once, then the state of the dyon reads
\begin{align}
\ket{\psi}= \frac{1}{\sqrt{2}}\bigg[\ket{\psi^{0}_{L}}+ {\rm e}^{i\oint_{C} \gradv\beta\cdot  d\v x}\ket{\psi^{0}_{R}}\bigg],
\end{align}
where we have dropped an unimportant overall phase factor.  Using $\beta\!=\!\theta \phi/(2\pi)$ in the phase shift in Eq.~(16), we obtain an alternative representation of the angle $\theta$ given by
\begin{align}
\oint_{C}\gradv\beta \cdot d \v x=\theta.
\end{align}
Since $\gradv\beta\!=\!(q\v A\! +\! g\v C)/(\hbar c)$ in a vacuum state, as seen in Eq.~(11), then we can say that the angle $\theta$ is a vacuum manifestation of the dyon phase in Eq.~(5) characterised by the winding number $n=1$. It follows that the angle $\theta$ in Eq.~(17) should manifest in the interference shift on the second screen. Following the procedure in Refs.~\cite{23,24}, we can show that this interference shift is given by
\begin{align}
\Delta x=  \frac{L \lambda}{d}\bigg(\bar{\delta}_0 + \frac{\theta}{2\pi}\bigg),
\end{align}
where $\bar{\delta}_0=\delta_0/2\pi$, with $\delta_0$ being the phase angle due to the momentum of the dyon having the de Broglie wavelength $\lambda$, $L$ is the distance between the two screens, $d$ is the separation between the two slits on the first screen, and we have assumed the condition $L>>d$.

A second possible manifestation of the angle $\theta$ in a vacuum state characterised by the winding number $n=1$ is through its associated differential scattering cross section due to a scattered dyon in the $x$-$y$ plane outside a dual flux tube. In the limit where the radius of the dual flux tube tends to zero (a dual flux line or string), we can follow the procedure given in Ref.~\cite{10} and obtain the differential scattering cross section attributed by the angle $\theta$, which is given by
\begin{align}
\frac{d \sigma_{\theta}}{d\Omega}= \frac{\sin^2(\theta/2)}{2\pi k \sin^2(\phi/2)},
\end{align}
where $\sigma_{\theta}$ is the cross-section, $\Omega$ is the solid angle, and $k$ is the magnitude of the dyon's wave vector.

We should note that the potential realisation of the above outlined interference experiments to detect the angle $\theta$ faces at least two fundamental problems: (i) the existence of dyons, which are still undetected but whose experimental search has recently begun \cite{25}. (ii) the idealisation of an infinitely long dual flux tube (a long solenoid does not completely confine the fields). This problem could be addressed by replacing the idealised infinitely long dual flux tube with a ``less-idealised'' toroidal dual solenoid (we note that the most convincing evidence of the AB effect was done using a toroidal solenoid \cite{26} and not a long solenoid. The toroidal solenoid may be reasonably modelled by a closed flux line whose theoretical treatment has recently been discussed \cite{24})

It is interesting to note that the dyon phase in Eq.~(5) takes the trivial value $\delta_{D}=2\pi N$, with $N$ integer, in the hypothetical case that the fluxes of the dual flux tube satisfy presumable quantisation rules analogous to those in Eq.~(2) for the case of the dyon charges, i.e., the flux quantisation rules 
\begin{align}
\Phi_{\rm e}=n_{\Phi_{\rm e}}\Phi^{0}_{\rm e}+n_{\Phi_{\rm m}}\frac{\Phi^{0}_{\rm e}\theta}{2\pi},\quad  \Phi_{\rm m}=n_{\Phi_{\rm m}}\Phi^{0}_{\rm m},
\end{align}
where $n_{\Phi_{\rm e}}$ and $n_{\Phi_{\rm m}}$ are integers and $\Phi^{0}_{\rm m}$ and $\Phi^{0}_{\rm e}$ are the flux quanta. In fact, if we use Eq.~(20) in Eq.~(5) it follows $\delta_{D}=2\pi N$ with $N=n(n_qn_{\Phi_{\rm m}}-n_gn_{\Phi_{\rm e}})$. However, following Witten's method \cite{5} that led to the electric charge quantisation in Eq.~(2), we can show that Eq.~(20) cannot be established for the fluxes in Eq.~(5). Witten's method demands two basic conditions: (i) the time-component of the Noether current associated with an infinitesimal rotation in the Lagrangian (in particular, Witten used the Lagrangian of the Georgi-Glashow model plus a $\theta$-term) must be non-vanishing and (ii) the fields in the Lagrangian must satisfy $\oint_S \v B\cdot d \v S \neq 0$ and $\oint_S \v E\cdot d \v S \neq 0$ at spatial infinity \cite{8}. When this method is applied to the fluxes in Eq.~(5) neither the condition (i) nor (ii) are fulfilled. The time-component of the corresponding Noether current vanishes because the Lagrangian does not depend on time derivatives of gauge potentials and the fields satisfy $\oint_S \v B\cdot d \v S = 0$ and $\oint_S \v E\cdot d \v S = 0$ at spatial infinity (there are no free charges in a dual flux tube \cite{17}). Thus, the quantisation rules in Eq.~(20) do not hold for the fluxes in the dual flux tube. We conclude that following Witten's approach, the angle $\theta$ affects the charges of the dyon but not the fluxes in the dual flux tube.

Finally, let us comment on the value $\theta\!=\!\pi$ associated with the only non-zero CP preserving angle. Recent experimental bounds yield \cite{27} $\theta < 1.98\times  10^{-10}$, indicating that CP is weakly violated. However, $\theta\!=\!\pi$ has applications in topological insulators \cite{28,29,30,31}. But in this case, the angle $\theta$ is not a universal constant but a quantity characterising the space filled by the insulator. In such configurations, a version of the Witten effect is realised by considering the concept of ``emergent monopoles.'' Hence, our results for $\theta\!=\!\pi$ could have some applicability in topological insulators.

Summarising. We have demonstrated that the dyon phase and the Witten effect imply the $\theta$-phase, which is topological, connected with CP violation, and quantised in a vacuum state: $\delta_{\theta}=n\theta$. We have shown that the most general vacuum state corresponding with the quantised $\theta$-phase is an Abelian form of the $\theta$-vacua. We have discussed two hypothetical interference effects in the vacuum where the angle $\theta$ could manifest. We have argued that the angle $\theta$ affects the dyon charges but not the fluxes of the dual flux tube. Although dyons are still unobserved, the possible detection of the $\theta$-phase would provide indirect evidence of these particles.

\end{document}